\documentclass[preprint,superscriptaddress,amsmath,showpacs,byrevtex,prb]{revtex4} 
\usepackage{graphicx}
\newcommand{\mib}[1]{\mbox{\boldmath$#1$}} 

\begin{document}
\title{Stochastic unraveling of Redfield master equations \\
  and its application to electron transfer problems}
\author{Ivan Kondov}
\altaffiliation[Present address:\ ]{Theoretische Chemie, Technische Universit\"at M\"unchen, 85747 Garching, Germany}
\affiliation{Institut f\"ur Physik, Technische Universit\"at, 09107 Chemnitz,
 Germany}
\author{Ulrich Kleinekath\"ofer}
\affiliation{International University Bremen, P.O.Box 750 561, 28725
  Bremen, Germany}
\author{Michael Schreiber}
\affiliation{Institut f\"ur Physik, Technische Universit\"at, 09107 Chemnitz,
 Germany}
\begin{abstract}
  A method for stochastic unraveling of general time-local quantum master
  equations (QMEs) is proposed. The present kind of jump algorithm allows a
  numerically efficient treatment of QMEs which are not in Lindblad form,
  i.e.\ are not positive semidefinite by definition. The unraveling can be
  achieved by allowing for trajectories with negative weights. Such a
  property is necessary, e.g.\ to unravel the Redfield QME and to treat
  various related problems with high numerical efficiency. The method is
  successfully tested on the damped harmonic oscillator and on electron
  transfer models including one and two reaction coordinates. The obtained
  results are compared to those from a direct propagation of the reduced
  density matrix (RDM) as well as from the standard quantum jump method.
  Comparison of the numerical efficiency is performed considering both the
  population dynamics and the  RDM in the Wigner phase space
  representation.
\end{abstract}

\pacs{03.65.Yz, 42.50.Lc, 34.70.+e, 82.20.-w}

\maketitle

\section{Introduction}
\label{sec:intro}

Time-independent as well as time-dependent phenomena in chemical physics,
quantum optics, solid state physics, biological physics, etc.\ are often
described using QMEs \cite{weis99,may00}. In particular, electron transfer
(ET) dynamics in open quantum systems, i.e., systems with dissipation, can
be conveniently treated within this formalism. The QMEs govern the time
evolution of density matrices which are used in order to represent the
mixed nature of the states.  Recently stochastic wave function methods have
received a great deal of attention. In unraveling schemes one considers an
ensemble of stochastic Schr\"odinger equations (SSEs) which in the limit of a
large ensemble resemble the respective QME.  Although all SSE approaches
have a common basis \cite{dors00a} they are usually divided into two
classes. One is the quantum jump method also known as Monte Carlo wave
function (MCWF) approach \cite{dali92,molm93,gard92,carm93,stei95}.  In this
approach the dynamics is described by a Schr\"odinger-like wave equation
interrupted by instantaneous deviations from the continuous motion (quantum
jumps). The second class of SSE approaches are the quantum diffusion models
with continuous motion \cite{gisi92,plen98} which are not in the center of
interest here. The numerical effort for solving SSEs scales much more
favorably with the size of the basis than a direct propagation of a density
matrix since one is dealing with wave functions and not density matrices
(for a comparison of direct integrators, see Ref.~\onlinecite{kond01}). Thus,
stochastic unraveling is an efficient numerical tool for solving QMEs. Of
course, to achieve good statistics one has to average over a large number
of wave functions. So the SSE approaches become preferable for large and
complex systems with many degrees of freedom. In passing, we want to
mention that in the present paper we are not interested in establishing a
relation between the SSE dynamics and some measurement process. Thus, the
treatment of single trajectories will be done without giving a special
physical meaning to them.

One of the important properties of a density matrix is the positive
semi-definiteness for all times, i.e.\ that all populations are positive or
zero. This property is fulfilled for QMEs of the Lindblad form
\cite{lind75} but not necessarily for reduced dynamics in general
\cite{pech94,gasp99a}. A slightly generalized generator for a completely
positive density-matrix evolution can be found in Ref.~\onlinecite{adle00}
while a discussion on the non-Markovian case has been done in
Ref.~\onlinecite{wilk00}.  Most of the unraveling schemes
\cite{dali92,gard92,gisi92,garr94,wolf95,plen98} have been restricted to
QMEs of Lindblad form \cite{lind75} which ensures that the RDM stays
positive semidefinite for all times and all parameters. Nevertheless, there
are many physically meaningful QMEs which result in positive semidefinite
RDMs although they are not of Lindblad form \cite{stru99,haak85}. The
increasing interest in descriptions beyond the Lindblad class such as the
quantum Brownian motion \cite{stru99,stru01}, the Redfield formalism
\cite{poll96,may00}, non-Markovian schemes \cite{cape94b,meie99,breu99},
etc.\ resulted in various efforts to develop new stochastic wave function
algorithms.

Strunz \emph{et al.}~\cite{stru99,stru01} extended the QME for Brownian
motion to a non-Markovian QME and then applied a quantum state diffusion
algorithm. A similar approach was also proposed by Gaspard \emph{et
  al.}~\cite{gasp99b}. Recently Stockburger and Grabert \cite{stoc02}
developed a method for an exact formulation of the RDM in terms of SSEs of
a system coupled to a linear heat bath. Breuer \emph{et al.}~\cite{breu99}
extended a scheme which they had used to calculate multi-time correlation
functions \cite{breu97} to the unraveling of QMEs. Their technique is based
on doubling the Hilbert space.  So instead of a single stochastic wave
function one has a pair of them \cite{breu99}. In this approach, norm and
Hermiticity are not preserved in single realizations but only in the
ensemble average which makes the algorithm unstable. Since stability and
efficiency are crucial issues for unraveling algorithms we propose an
alternative approach which fulfills these criteria.  Though the present
approach has only been tested for Redfield and Brownian-dynamics master
equations so far \cite{klei02} there are only few restrictions to its range
of validity and it is therefore applicable to a much larger class of
time-local quantum master equations.

ET is commonly treated in modern theories with use of the RDM formalism and
QMEs. Alternatives to QMEs are, for example, semi-classical theories
\cite{goyc01}, path integral methods \cite{makr96,sim97} and recently the
self-consistent hybrid approach \cite{thos01,thos02,wang03}. The latter was
shown to treat successfully the spin-boson problem \cite{thos01}, the ET in
mixed-valence compounds \cite{thos02,wang03} as well as the heterogeneous
ET at semiconductor surfaces \cite{wang03}. Solving the QME for ET in model
systems with one \cite{may92,jean92,felt95,fuch96} and many reaction
coordinates \cite{wolf96} has been done with success. Exhaustive reviews on
ET can be found, e.g., in Refs.~\onlinecite{barb96} and
\onlinecite{bixo99}. Apart from the non-Markovian descriptions of transfer
phenomena \cite{cape94a,cape94b,cape95,manc01c} the use of Redfield theory
for ET was investigated as well
\cite{jean92,felt95,kond01,egor01,klei01,schr01,egor03}. The model used in
the latter references is based on vibronically coupled diabatic potentials
which are sufficiently well approximated by harmonic potentials. In
particular, the influence of the electronic coupling between the diabatic
states on the dissipation was investigated. Neglecting this effect results
in the diabatic damping approximation (DDA) \cite{egor01,klei01}. This
approximation as well as considering first order perturbation theory in the
electronic coupling were objects of recent studies \cite{klei01}. A typical
problem that occurs with increasing the complexity of the ET models, i.e.\ 
the dimension of the RDM, is the numerical effort. Thus, the stochastic
unraveling of generalized time-local QME was developed with the prospect of
applications to more complex ET systems.

Recently the present scheme was briefly demonstrated for the quantum
Brownian motion of a harmonic oscillator \cite{klei02}. In the present
paper the stochastic unraveling of the Redfield QME shall be considered in
more detail as well as applications concerning multi-mode models for ET
shall be presented.  In the next Section a brief introduction to the
Redfield formalism will be given. Section~\ref{sec:sse} focuses on the
derivation of the SSEs relevant for the generalized time-local QME while
Section \ref{sec:jump_rates} will provide explicit expressions for the jump
rates. In Section~\ref{sec:app} we describe three concrete applications of
the proposed quantum jump method: the damped harmonic oscillator and a
model for ET with one and two reaction modes. A study and discussion of the
numerical efficiency in Section~\ref{sec:efficiency} and a conclusion
follow. The detailed quantum jump algorithm used in the present
contribution can be found in the Appendix.  Atomic units are used
throughout the paper, i.e.\ $\hbar=1$.

\section{Redfield formalism}
\label{sec:redfield}

In Redfield theory the overall system is partitioned into a relevant system
whose evolution is of interest and a thermal bath using the Hamiltonian
\begin{equation}
H = H_{\rm S} + H_{\rm B} + H_{\rm SB}.
\label{eq:Hamiltonian}
\end{equation}
Here $H_{\rm S}$ and $H_{\rm B}$ are the Hamiltonians of the relevant
system and of the bath while $H_{\rm SB}$ describes their interaction. In
Subsec.~\ref{subsec:et-2d} it will be shown how this partitioning can be
rigorously performed. In general, the interaction part can be represented
by bilinear products
\begin{eqnarray}
H_{\rm SB}=\sum\limits_m\Phi_m K_m
\label{eq:sb}
\end{eqnarray}
of system and bath operators, $K_m$ and $\Phi_m$, respectively. In the
following $K_m$ and $\Phi_m$ will be considered Hermitian.  The state of
the system is described by the RDM performing a trace of the total density
matrix $\sigma$ over the bath degrees of freedom, i.e.\ $\rho={\rm tr}_{\rm
  B} \sigma$. It is assumed that the bath stays in thermodynamic
equilibrium at all times. This means that the relaxation of the bath is
much faster than the evolution of the system.  In addition, one assumes
that the system-bath interaction is sufficiently small to be treated
perturbatively to second order. Using the Hamiltonian
(\ref{eq:Hamiltonian}) and the assumptions described above one obtains a
non-Markovian QME for the RDM. One possible way to obtain a Markovian QME
instead is to neglect memory effects which are due to the finite bath
correlation time. The formal treatment yields the Redfield QME
\cite{poll94,may00}
\begin{eqnarray}
\dot{\rho} = - i \left[ H_{\rm S},\rho \right]
+ \sum\limits_m
\bigg\{
\left[ \Lambda_m\rho,K_m\phantom{^\dagger} \right]+
\left[ K_m,\rho \Lambda_m^{\dagger} \right]
\bigg\}~.
\label{eq:pf-form}
\end{eqnarray}
Note that the operator $\Lambda_m^{\dagger}$ is the adjoint of the
relaxation operator $\Lambda_m$. This is only true if $K_m$ and $\Phi_m$
are Hermitian \cite{gasp99a,may00} but not in general \cite{poll94,poll96}.
The relaxation operator is given by
\begin{equation}
\Lambda_m=\sum\limits_n\int\limits_0^{\infty}d\tau C_{mn}(\tau)
 K_n^{\rm{I}}(-\tau),
\label{eq:lambda}
\end{equation}
where $C_{mn}(\tau)$ is the bath correlation function and $K_n^{\rm{I}}$
the system operator in the interaction picture. Usually it is easier to
obtain the latter quantity in the frequency domain, e.g.\ with use of
molecular dynamics simulations or with a simple bath modeling. Either
approach yields the bath spectral density $J_{mn}(\omega)$ in terms of
which the correlation function can be constructed as \cite{may00}
\begin{eqnarray}
\label{eq:corr}
  C_{mn}(\omega)=2 \pi [1+n(\omega)][J_{mn}(\omega)-J_{mn}(-\omega)]
\end{eqnarray}
with the Bose-Einstein distribution $n(\omega)=(e^{\beta\omega}-1)^{-1}$
and $\beta=(k_{\rm B}T)^{-1}$ being the inverse temperature. All
considerations in the present work will be limited to the Ohmic form of the
spectral density with exponential cut-off. However, a spectral density of
Debye form can be constructed which results in nearly the same values of
$J_{mn}(\omega)$ for the specific spectrum of $H_{\rm S}$ used.

Under certain approximations Eq.~(\ref{eq:pf-form}) can be transformed to
Lindblad form \cite{egor01}
\begin{equation}
\label{eq:lindblad-form}
\frac{d\rho(t)}{d t}=
-i
\left[H_{\rm S},\rho(t)\right]
+\sum\limits_n
\left[
L_n\rho(t)L_n^\dagger
-\frac{1}{2}\rho(t)L_n^\dagger L_n
-\frac{1}{2}L_n^\dagger L_n\rho(t)
\right].
\end{equation}
One way to obtain the Lindblad QME (\ref{eq:lindblad-form}) is starting
either from the non-Markovian QME or from Eq.~(\ref{eq:lambda}) and
assuming a $\delta$-correlated bath\cite{gasp99a,may00}, i.e., $C_{mn}(\tau)\rightarrow
c_{mn}\delta(\tau)$. Subsequent diagonalization of the
correlation matrix $\mib{c}$ by means of a unitary transformation
$\mib{c}=U^\dagger \mib{\kappa} U$, yields Eq.~(\ref{eq:lindblad-form})
with $L_n=\sqrt{\kappa_n}\sum_m U_{nm}K_m$. Alternatively, the Lindblad QME
is obtained when the DDA \cite{egor01,klei01} is invoked with the so-called
rotating wave approximation (RWA) in the system-bath coupling as done in
the present paper. The explicit form of the corresponding Lindblad
operators used will be given in the sections below.

The Lindblad QME (\ref{eq:lindblad-form}) will be used to compare the
numerical efficiency of the present quantum jump method method with the
standard one \cite{dali92,gard92,gisi92,garr94,plen98} by solving the
single-mode and the two-mode ET model. For more details of the DDA we refer
to Refs.~\onlinecite{egor01} and \onlinecite{klei01}.

\section{Stochastic Schr\"odinger equation}
\label{sec:sse}

To start with the derivation of the unraveling scheme we first like to
formulate the time-local QME in its most general form
\begin{eqnarray}
\frac{d\rho(t)}{dt} &=& A(t)\rho(t)+\rho(t)A^\dagger(t) \nonumber \\
&&+\sum\limits_{k=1}^M \big\{C_k(t)\rho(t) E_k^\dagger(t)
+E_k(t)\rho(t)C_k^\dagger(t)\big\}~.
\label{eq:qme}
\end{eqnarray}
Because $A(t)$, $C_k(t)$, and $E_k(t)$ are arbitrary operators this
equation conserves only Hermiticity. In order to conserve also the norm
further restrictions have to be applied as shown below.  All time arguments
will be dropped henceforth for clarity.

The RDM will be recovered by averaging over an ensemble of two vectors
$|\psi\rangle$ and $|\phi\rangle$, which are elements of the doubled
Hilbert space, as
\begin{eqnarray}
\rho = \overline{
 |\psi\rangle\langle\phi|}+\overline{|\phi\rangle\langle\psi|}~.
\label{eq:recon}
\end{eqnarray}
Every individual realization of the stochastic process before averaging
denoted by the pair $\left(|\psi\rangle,|\phi\rangle\right)$ will be called
a trajectory. In contrast to Ref.~\onlinecite{breu99} the averaging formula
(\ref{eq:recon}) preserves Hermiticity of single trajectories leading to a
significantly improved numerical performance of the scheme. Each trajectory
$\left(|\psi\rangle,|\phi\rangle\right)$ is propagated be means of two SSEs
having the following generic form
\begin{subequations}
\label{eq:sse0}
\begin{eqnarray}
d|\psi\rangle&=&
D_1|\psi\rangle dt
+\sum\limits_{k=1}^M\sum\limits_{i=1}^2
 S_{1k}^i|\psi\rangle d \xi_k^i~,
\label{eq:sse0_a}
\\
d|\phi\rangle&=&
D_2|\phi\rangle dt
+\sum\limits_{k=1}^M\sum\limits_{i=1}^2
 S_{2k}^i|\phi\rangle d \xi_k^i~.
\label{eq:sse0_b}
\end{eqnarray}
\end{subequations}
Unlike deterministic differential equations the SSEs include differentials
of the complex noise variables $\xi_k^i$ in addition to the time variable.
The superscript in $\xi_k^i$ denotes which of the two terms from the
Hermitian pair in the sum in Eq.~(\ref{eq:qme}) is taken while the
subscript counts the relevant dissipative channels from $1$ to $M$. The
operators $D_1$ and $D_2$ specify the deterministic and the operators
$S_{jk}^i$ the stochastic part of the evolution. In general, they may be
time-dependent. The stochastic differentials\cite{gard85} $d\xi^i_k$ are
assumed to have zero mean, to be uncorrelated and normalized to $dt$:
\begin{eqnarray}
\overline{d\xi^i_k}=0,\ \overline{d\xi^{i\ast}_k d\xi^j_{l}}=
\delta_{ij}\delta_{kl}dt~.
\label{eq:noise}
\end{eqnarray}
Differentiating Eq.~(\ref{eq:recon}), neglecting all terms higher than
first order in $dt$, and assuming that ensemble averages always factorize
\cite{dors00a} yields
\begin{eqnarray}
d\rho&=&
\left[
 D_1 \overline{|\psi\rangle\langle\phi|}
+D_2 \overline{|\phi\rangle\langle\psi|}
\right] dt \nonumber \\
&&+\sum\limits_{k=1}^M
\left[
 S_{1k}^1|\overline{\psi\rangle\langle\phi|}S_{2k}^{1\dagger}
+S_{2k}^2|\overline{\phi\rangle\langle\psi|}S_{1k}^{2\dagger}
\right] dt
+
\textrm{H.c.}
\label{eq:qme_sub}
\end{eqnarray}
Comparing Eq.~(\ref{eq:qme_sub}) with the original
QME~(\ref{eq:qme}) one is able to replace
\begin{eqnarray}
S_{1k}^2=S_{2k}^1=C_k+\alpha_k^1
\quad\textrm{and}\quad
S_{1k}^1=S_{2k}^2=E_k+\alpha_k^2
\label{eq:s-constr}
\end{eqnarray}
where $\alpha_k^1$ and $\alpha_k^2$ are arbitrary possibly time-dependent
scalar functions of $\left(|\psi\rangle,|\phi\rangle\right)$. Plugging the latter expressions into Eq.~(\ref{eq:qme_sub}) yields
\begin{eqnarray}
D_1=D_2=A-\sum\limits_{k=1}^M
\left(
\alpha_k^{2\ast} C_k +\alpha_k^{1\ast} E_k + \alpha_k^1\alpha_k^{2\ast}
\right).
\label{eq:d-constr}
\end{eqnarray}
According to Ref.~\onlinecite{dors00a} Eq.~(\ref{eq:sse0}) describes a quantum
diffusion process if the leading terms in $d\xi^i_k$ are of first order in
$\sqrt{dt}$. When $d\xi^i_k$ can be given by a finite number of values
only, e.g.\ $\pm{}\sqrt{dt}$, the process results in continuous but random
trajectories within each infinitesimal time interval $dt$ (for
$dt\rightarrow 0$ the trajectories become smooth but still stay noisy). In
that way one derives diffusion methods which will not be considered in the
present work. However, if the leading terms in $d\xi^i_k$ have finite
values of order unity, i.e.\ zeroth order in $\sqrt{dt}$,
Eq.~(\ref{eq:sse0}) leads to the so-called quantum jump methods which
produce trajectories that are deterministic during finite time intervals
connected by discontinuous transitions (jumps). The jumps are specified by
their jump rates $p^i_k$, which have to be real scalar functions of
$\left(|\psi\rangle,|\phi\rangle\right)$. If $n_k^i(t)$ is the number of
jumps in channel $k$ due to term $i$ up to time $t$, the probability for
$n_k^i(t)$ to increase by one, i.e.\ the expectation value of both $dn_k^i$
and $(dn_k^i)^2$, should be equal to $p^i_k dt$ during the infinitesimal
time interval $dt$. This can be written as \cite{dors00a}
\begin{eqnarray}
d\xi^i_k=\frac{dn^i_k - p^i_k dt}{\sqrt{p^i_k}}e^{i\varphi}
\label{eq:noise-jumprate}
\end{eqnarray}
so that it obeys condition (\ref{eq:noise}). The phase factor
$e^{i\varphi}$ leads merely to a phase shift in the wave vectors and
cancels within each realization and we therefore set $\varphi=0$. If
$dn_k^i$ vanishes for all $k$ and $i$, then Eq.~(\ref{eq:sse0}) becomes a deterministic Schr\"odinger equation. For any $k$ and $i$, $dn_k^i=1$
indicates the occurrence of a jump. In this case we have
$d(|\phi\rangle,|\psi\rangle)=(|\phi\rangle,|\psi\rangle)_\textrm{after
  jump}-(|\phi\rangle,|\psi\rangle)_\textrm{before jump}$. Taking this into
account and substituting Eqs.~(\ref{eq:noise-jumprate}) and
(\ref{eq:s-constr}) into Eq.~(\ref{eq:sse0}) it is found that
$\alpha_k^i=-\sqrt{p^i_k}$. Eventually, the final form of the SSEs for the
quantum jump method is obtained as
\begin{subequations}
\label{eq:sse1}
\begin{eqnarray}
\label{eq:sse1_a}
d|\psi\rangle&=&
\left(
A+\sum\limits_{k=1}^M\frac{p_k^1+p_k^2}{2}
\right)
|\psi\rangle dt
 \\
&&+\sum\limits_{k=1}^M
\left[ 
  \left( \frac{E_k}{\sqrt{p_k^1}}-1 \right) dn_k^1
+ \left( \frac{C_k}{\sqrt{p_k^2}}-1 \right) dn_k^2
\right]
|\psi\rangle, \nonumber
\\
\label{eq:sse1_b}
d|\phi\rangle&=&
\left(
A+\sum\limits_{k=1}^M\frac{p_k^1+p_k^2}{2}
\right)
|\phi\rangle dt
 \\
&&+\sum\limits_{k=1}^M
\left[ 
  \left( \frac{C_k}{\sqrt{p_k^1}}-1 \right) dn_k^1
+ \left( \frac{E_k}{\sqrt{p_k^2}}-1 \right) dn_k^2
\right]
|\phi\rangle. \nonumber
\end{eqnarray}
\end{subequations}

\section{Jump rates}
\label{sec:jump_rates}
Essential for the performance and particularly for the convergence behavior
of the quantum jump method is how the jump rates $p_k^1$ and $p_k^2$ are
specified. They have no physical meaning since in the average they do not
influence any observable but determine the statistical error. A detailed
discussion on the optimization of the jump rates as free parameters can be
found in Ref.~\onlinecite{felb99}. Another freedom is that
Eq.~(\ref{eq:qme}) is invariant with respect to a gauge transformation of
the kind $C_k\rightarrow f C_k,\ E_k\rightarrow E_k/f$ if $f$ is a real,
scalar function of time. Each single realization, and hence the stochastic
process, is independent of this gauge transformation and using such
transformation offers us no further advantages. We note that the jump rates
in Ref.~\onlinecite{breu99} do not fulfill the invariance under this gauge
transformation.

Following the approach in Ref.~\onlinecite{felb99} we require that the norm of
every single trajectory is constant in time. Under such a condition
expressed as
\begin{eqnarray}
\textrm{tr}\left\{\frac{d}{dt}\left[|\psi\rangle\langle\phi|+
|\phi\rangle\langle\psi|\right]\right\}=0
\label{eq:nconserve}
\end{eqnarray}
the $p_k^i$ are adapted at each moment of time. This approach yields a
numerically stable and efficient algorithm.  In contrast, numerical tests
with jump rates adapted to other quantities such as $\langle \phi|\phi
\rangle$, $\langle \psi|\psi \rangle$, etc.\ resulted in an unstable
scheme. The operators that enter the QME~(\ref{eq:qme}) are restricted by
condition (\ref{eq:nconserve}) yielding
\begin{eqnarray}
\label{eq:a-c-e}
A+A^\dagger
+\sum_{k=1}^M 
\left(
 E^\dagger_k C^{\phantom\dagger}_k
+C^\dagger_k E^{\phantom\dagger}_k
\right)
=0.
\end{eqnarray}
Let us try to determine the jump rates from this condition.  The total jump
rate is obtained applying Eq.~(\ref{eq:nconserve}) to the deterministic
part of Eq.~(\ref{eq:sse1}):
\begin{eqnarray}
\label{eq:tot-rate_tilde}
\tilde{p}=-\frac
{\langle\phi|A+A^\dagger|\psi\rangle
+\langle\psi|A+A^\dagger|\phi\rangle}
{\langle\phi|\psi\rangle+\langle\psi|\phi\rangle}.
\end{eqnarray}
All partial jump rates can be then successively found using
Eqs.~(\ref{eq:a-c-e}) and (\ref{eq:tot-rate_tilde}):
\begin{subequations}
\label{eq:jump-rates_tilde}
\begin{eqnarray}
\tilde{p}_k^1&=&\frac{
 \langle\phi|C_k^\dagger E^{\phantom\dagger}_k|\psi\rangle
+\langle\psi|E_k^\dagger C^{\phantom\dagger}_k|\phi\rangle}
{\langle\phi|\psi\rangle+\langle\psi|\phi\rangle},
\label{eq:jump-rates_a_tilde}
\\
\tilde{p}_k^2&=&\frac{
 \langle\phi|E_k^\dagger C^{\phantom\dagger}_k|\psi\rangle
+\langle\psi|C_k^\dagger E^{\phantom\dagger}_k|\phi\rangle}
{\langle\phi|\psi\rangle+\langle\psi|\phi\rangle}.
\label{eq:jump-rates_b_tilde}
\end{eqnarray}
\end{subequations}
Here a problem occurs because the values of the  $\tilde{p}_k^i$ do not
have to be positive for all trajectories at all times. But, since the jump
rates $p_k^i$ are arbitrary real functions we can choose them as the
absolute values of the $\tilde{p}_k^i$
\begin{subequations}
\label{eq:jump-rates}
\begin{eqnarray}
p_k^1&=& \left| \frac{
 \langle\phi|C_k^\dagger E^{\phantom\dagger}_k|\psi\rangle
+\langle\psi|E_k^\dagger C^{\phantom\dagger}_k|\phi\rangle}
{\langle\phi|\psi\rangle+\langle\psi|\phi\rangle} \right|,
\label{eq:jump-rates_a}
\\
p_k^2&=& \left| \frac{
 \langle\phi|E_k^\dagger C^{\phantom\dagger}_k|\psi\rangle
+\langle\psi|C_k^\dagger E^{\phantom\dagger}_k|\phi\rangle}
{\langle\phi|\psi\rangle+\langle\psi|\phi\rangle} \right| .
\label{eq:jump-rates_b}
\end{eqnarray}
\end{subequations}
An additional weight factor $\pm{}1$ for the trajectories has to be introduced
which changes its sign every time a jump is performed with
$p_k^i=-\tilde{p}_k^i$. It can also be implemented (as in the appendix) by
allowing for negative norms of the trajectories. The change from
$\tilde{p}_k^i$ to $p_k^i$ gives rise to a small deviation of the norm from
unity because in the regions where not all $\tilde{p}_k^i$ and $p_k^i$ are
identical, norm conservation is no longer guaranteed, i.e. the sum of the
$p_k^i$ differs from $\tilde{p}$. As long as the occurrence of a jump is a
very rare event and the number of negative $\tilde{p}_k^i$ is also very
small the deviation from the initial norm is expected to be small. In all
tests this deviation was far below 1\% without effecting the numerical
efficiency of the proposed algorithm.  This is how the scheme tolerates
trajectories with possibly negative weights which arise from the fact that
the RDM with the QME~(\ref{eq:qme}) is not necessarily positive
semidefinite. If the RDM stays positive semidefinite during its entire time
evolution the negative weights are not needed, i.e.\ all trajectories can
be normalized to unity and represent physically relevant pure states. A
possible implementation of the present unraveling scheme is shown in the
appendix.

In the examples below the RDM can exhibit negative eigenvalues. This
nonphysical situation could probably be improved by applying an initial
slippage to the initial state \cite{gasp99a}. 
Another possibility to avoid
non-positive semidefinite RDMs is to start with a derivation of
different QMEs in the form (\ref{eq:qme}) with time-dependent coefficients.
It has been shown that non-Lindblad QMEs with time-dependent coefficients
can preserve the positivity of the RDM \cite{stru99,haak85}. Nevertheless,
an unraveling scheme has to be able to follow also the nonphysical behavior of the QME because in the ensemble average the solution of the SSEs should completely coincide with the exact solution of the QME.

\section{Applications of the present unraveling method}
\label{sec:app}

\subsection{Damped harmonic oscillator}
\label{subsec:qho}

One of the most simple toy models used for testing in dissipative quantum
dynamics is the damped harmonic oscillator. Here it will be formulated
within Redfield theory, i.e.\ one has to obtain the explicit form of the
operators in Eq.~(\ref{eq:pf-form}). The oscillator has mass ${\cal M}$ and
frequency $\omega_0$. If the thermal bath is modeled by quantum harmonic
oscillators and the system operator $K$ is the oscillator coordinate
$q=(a+a^\dagger)/\sqrt{2{\cal M}\omega_0}$ the sum in
Eq.~(\ref{eq:pf-form}) contains only one term in which
\begin{eqnarray}
\label{eq:lambda-ho}
\Lambda
=
\Gamma\sqrt{\frac{{\cal M}\omega_0}{2}}
\left[
\left(
n(\omega_0)+1
\right)a
+
n(\omega_0)a^\dagger
\right].
\end{eqnarray}
The damping rate $\Gamma$ is related to the spectral density of the bath
$J(\omega)$ as $\Gamma=\pi J(\omega_0)/({\cal M}\omega_0)$. Therefore, the
explicit form of $J(\omega)$ is not necessary since the oscillator and the
bath interchange quanta only at the frequency $\omega_0$. Performing either the
RWA \cite{wolf97dis,kohe97b} or the secular approximation
\cite{may00,lind00,kohe97b} Eq.~(\ref{eq:pf-form}) is transformed into a
Lindblad QME (\ref{eq:lindblad-form}). For the sake of simplicity this will
be shown here with the RWA for the harmonic oscillator but the generalization
for the ET model solved within the DDA is straightforward. Inserting the
expressions for $\Lambda$ and $K$ into
Eq.~(\ref{eq:pf-form}), denoting $b_1=a$ and $b_2=a^\dagger$, and
performing some calculus the QME obtains the form
\begin{eqnarray}
\label{eq:qme-pre-rwa}
\dot{\rho} = - i \left[ H_{\rm S},\rho \right]
+
\frac{\Gamma}{2}\sum\limits_{i,j=1}^2
\kappa_{ij}
\left(
b_i\rho b_j^{\dagger}-\frac12 b_j^{\dagger}b_i\rho-\frac12 \rho b_j^{\dagger}b_i
\right),
\end{eqnarray}
where $\mib{\kappa}$ is the correlation matrix

\begin{eqnarray}
\label{eq:corr-matr}
\mib{\kappa}=
\left(
\begin{array}{cc}
n(\omega_0)             &       n(\omega_0)+\frac12\\
n(\omega_0)+\frac12     &       n(\omega_0)+1\\
\end{array}
\right).
\end{eqnarray}
In order to transform Eq.~(\ref{eq:qme-pre-rwa}) into Lindblad form either
$\mib{\kappa}$ has to be diagonalized imposing conditions for which the
eigenvalues are positive \cite{dum92b} or the RWA in the system-bath
coupling \cite{kohe97b} has to be performed. It is easily seen that the
determinant of $\mib{\kappa}$ is $-\frac14$ and hence the former method
fails for this QME. Performing the RWA implies that the off-diagonal
elements of $\mib{\kappa}$ are set to zero. Then the Lindblad operators
take the explicit form
\begin{equation}
L_1 = \sqrt{(n(\omega_0)+1)\Gamma}a
\quad \textrm{and}\quad 
L_2 = \sqrt{n(\omega_0)\Gamma}a^\dagger~.
\label{eq:lb-oper}
\end{equation}
This result is easily generalized for the ET models discussed in the next
two subsections. $L_1$ and $L_2$ have a clear physical interpretation.
$L_1$ damps all occupied levels bringing their populations one level lower,
while $L_2$ has the opposite effect. In the thermodynamic equilibrium the
jump rates for both operators are equal: the populations do not change.

To find the operators involved in the generalized QME~(\ref{eq:qme}) (with
$M=1$ for the harmonic oscillator) one has to carry out the commutators in
Eq.~(\ref{eq:pf-form}). Then one can easily identify
\begin{eqnarray}
C_1=K,\ E_1=\Lambda, \ A=-i H_{\rm S}-K\Lambda.
\label{eq:pf-form-op}
\end{eqnarray}
In contrast to the Lindblad operators (\ref{eq:lb-oper}) the action of
$C_1$ and $E_1$ on the wave function is more subtle. This is why it is
difficult, and probably not possible, to assign a
certain physical process to a single trajectory.

\begin{figure}
\includegraphics[width=8.5cm]{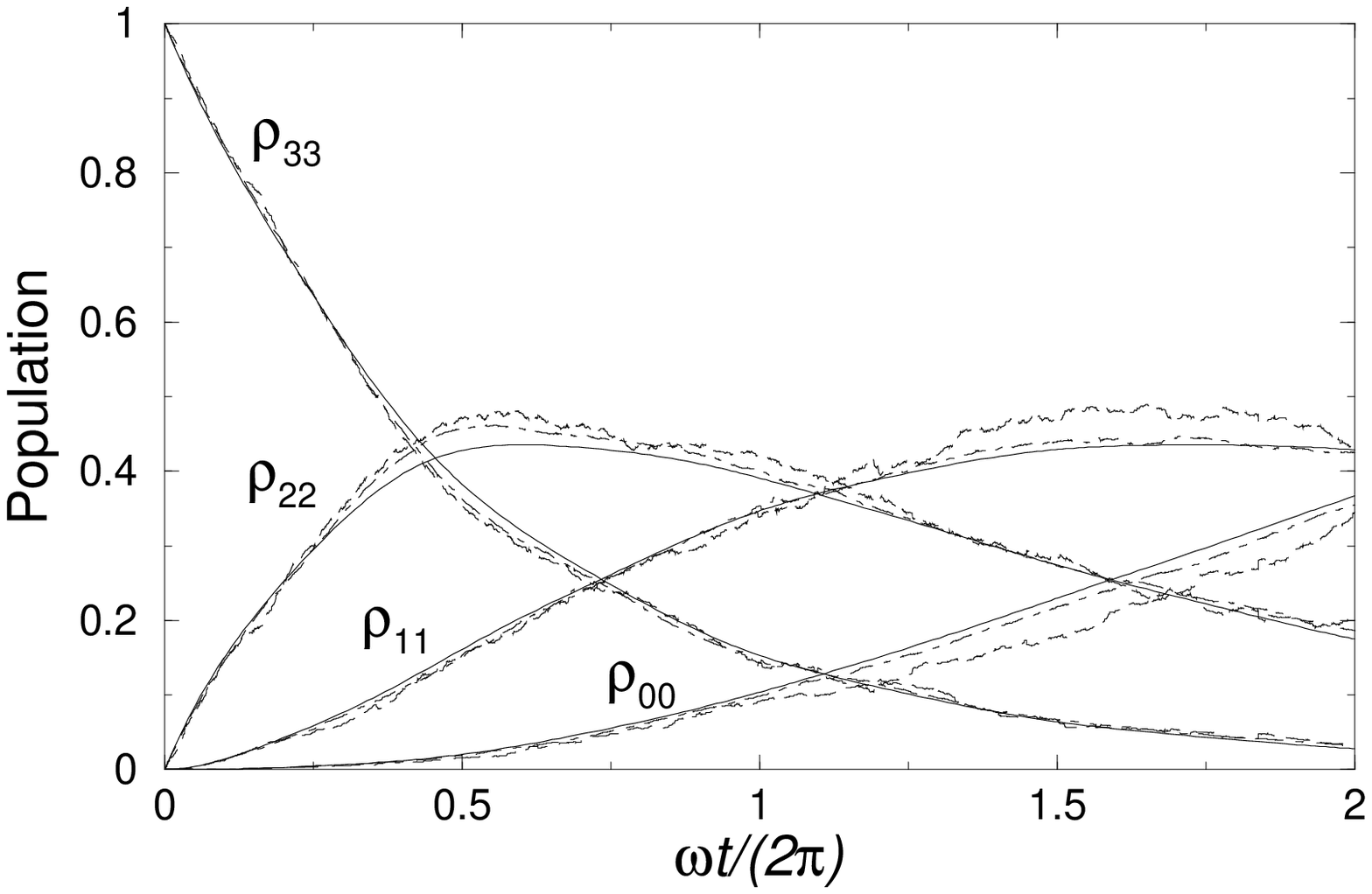}
\caption{
\label{fig:ho-pop}
Population dynamics of a damped harmonic oscillator simulated using the
proposed quantum jump method with $10^3$ trajectories (dashed lines) and
with $10^4$ trajectories (dot-dashed lines) compared to the direct RDM
propagation (solid lines).}
\end{figure}

In our stochastic simulation of the damped harmonic oscillator the
temperature $T=\omega_0/4$ and $\Gamma=\omega_0/10$ are used.
Figure~\ref{fig:ho-pop} shows the population dynamics of the lowest four
levels of the oscillator starting from the pure initial state
$\rho_{33}=1$. As seen, the convergence to the exact solution is very slow
($10^4$ trajectories are still not sufficient).  On the other hand the test
system is very small and the QME can be solved very fast using direct
propagators. The true advantage of the method can be seen with larger
systems, where it shows both a faster convergence and a good scaling.

\subsection{Electron transfer model with one reaction coordinate}
\label{subsec:et}
Let us consider a model for electron transfer with the Hamiltonian
\cite{klei01}
\begin{eqnarray}
H_{\rm S}=H^{(0)}+V=\sum\limits_i H_i |i\rangle\langle i|+V
\label{eq:h_s}
\end{eqnarray}
where $H_i$ are the Hamiltonians of two harmonic oscillators (i.e., 
$i=1,2$) which describe the vibronic spectrum of two electronic states interacting via the electronic coupling $V$. If the system includes a
single reaction coordinate $q$ the vibronic Hamiltonians read
\begin{eqnarray}
H_i=U_i
+\omega_i(a_i^\dagger a_i+\frac12)
+\frac{\omega_i\Delta_i}{\sqrt2}(a_i + a_i^\dagger),
\label{eq:h_0}
\end{eqnarray}
where $a_i$ and $a_i^\dagger$ are the boson operators, $\Delta_i$ the
dimensionless displacements of the harmonic potentials along the reaction
coordinate, $\omega_i$ the oscillator frequencies, and $U_i$ the electronic
excitation energies. A very useful parameter of the system which is related
to the last term in Eq.~(\ref{eq:h_0}) is the reorganization energy
$\lambda_i=\omega_i\Delta_i^2/2$. It is also proportional to the vibronic
coupling $\omega_i\Delta_i$.  Using the former expression one can define
the potential minima as $U^0_i=U_i-\lambda_i$.  Configurations in which the
potential minimum of the upper free-energy surface is lower in energy than
the lower free-energy surface at that point are in the so-called
\emph{normal region}.  If the opposite is true the configuration is in the
\emph{Marcus inverted region}. The electronic coupling $v_{12}$ between the
model potential surfaces is independent of the coordinate. So the
respective term in Eq.~(\ref{eq:h_s}) obtains the explicit form
\begin{eqnarray}
V=
\sum_{i,j} \sum_{M,N}
(1-\delta_{ij})
v_{ij} 
f(i,M;j,N) 
|iM\rangle \langle jN|~.
\label{eq:v}
\end{eqnarray}
The Franck-Condon factors $f(i,M;j,N)$ are calculated using the
eigenfunctions $\varphi_{iM}$ of the harmonic oscillators
\begin{equation}
f(i,M;j,N)
=
\langle iM|jN\rangle
=
\int dq \varphi_{iM}(q)\varphi_{jN}(q)~.
\label{eq:fcf}
\end{equation}
By analogy with the damped harmonic oscillator the system operator $K$ is defined as the coordinate operator, i.e.\ 
\begin{equation}
K = q = \sum\limits_i \left( 2\omega_i {\cal M} \right)^{-1/2} \left( a_i^{\dagger} +a_i \right) |i \rangle\langle i|~.
\label{eq:ka}
\end{equation}

We consider a potential configuration in the normal region with no barrier
between the two harmonic potentials which have equal curvature
($\omega_1=\omega_2\equiv\omega_0$), change of free energy
$U_2^{(0)}-U_1^{(0)}=-2\omega_0$, reorganization energy
$\lambda_1=3\omega_0$, $\lambda_2=0$ and inter-center coupling
$v_{12}=\omega_0$. The reason for this choice is the intention to compare
the standard and the new quantum jump methods. As the Lindblad QME is
obtainable only with RWA and DDA one has to study a parameter region where
both QMEs generate almost the same dynamics. The bath is described by an
Ohmic spectral density with cut-off frequency $\omega_c=\omega_0$ at
temperature $k_\textrm{B}T=\omega_0/4$. The system-bath interaction is
characterized by the damping rate $\Gamma=\pi\eta/({\cal
  M}\exp(1))=\omega_0/10$ (see Ref.~\onlinecite{klei01} for details).
Again, the operators necessary for the new quantum jump method are defined according to Eq.~(\ref{eq:pf-form-op}).

\begin{figure}
\includegraphics[width=8.5cm]{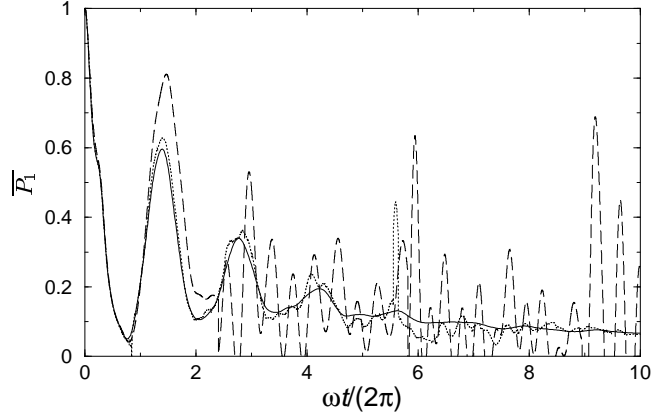}
\caption{
\label{fig:et-pop}
Relaxation of the donor population for the electron transfer model with a
single reaction mode. The solid line shows the exact solution of the QME,
the dashed line one arbitrary trajectory of the quantum jump method, the
dotted line an average over 500 trajectories.}
\end{figure}

\begin{figure}
\includegraphics[width=8.5cm]{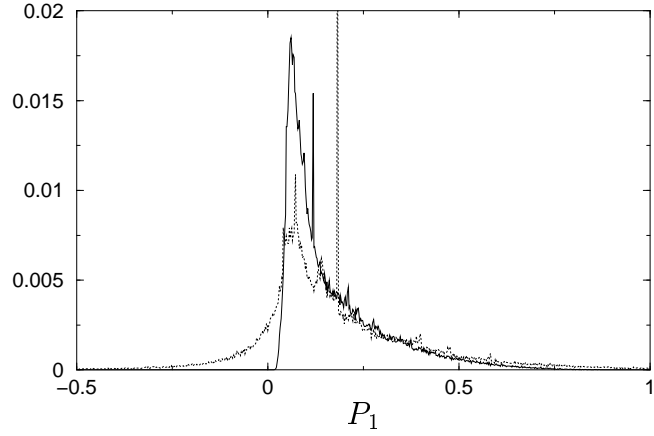}
\caption{
\label{fig:et-distrib}
Distribution of the expectation values of the population of the donor state
$P_1$ produced by the new unraveling scheme for the Redfield QME (dotted
line) and the standard normalized jump method for the Lindblad QME (solid
line) at time $\omega t/(2\pi)=5.8$. Both distributions are normalized to
unity. }
\end{figure}

A Gaussian wave packet located at the donor state $|1\rangle$ having energy
slightly above the crossing of the harmonic potentials was chosen as
initial state. The numerical simulation with about 1000 trajectories
provides sufficiently converged and accurate results.
Figure~\ref{fig:et-pop} shows the relaxation of the ensemble-averaged donor
population $\overline
{P_1}=\overline{\langle\psi|1\rangle\langle1|\phi\rangle}+
\overline{\langle\phi|1\rangle\langle1|\psi\rangle}$. A widely discussed
property of the Redfield equation is that it does not strictly conserve
positivity of the RDM \cite{kohe97b}. Although $\overline{P_1}$ is always
positive the tiny negative fraction in Fig.~\ref{fig:et-distrib} is an
evidence for the existence of single realizations with negative $P_1$. In
contrast, the simulation of the same system with the Lindblad operators
(\ref{eq:lb-oper}) by means of the standard quantum jump method
\cite{dali92,gard92,gisi92,garr94,plen98} keeps all values of $P_1$ well
confined between $0$ and $1$.

\begin{figure}
\includegraphics[width=12.5cm,angle=180,clip]{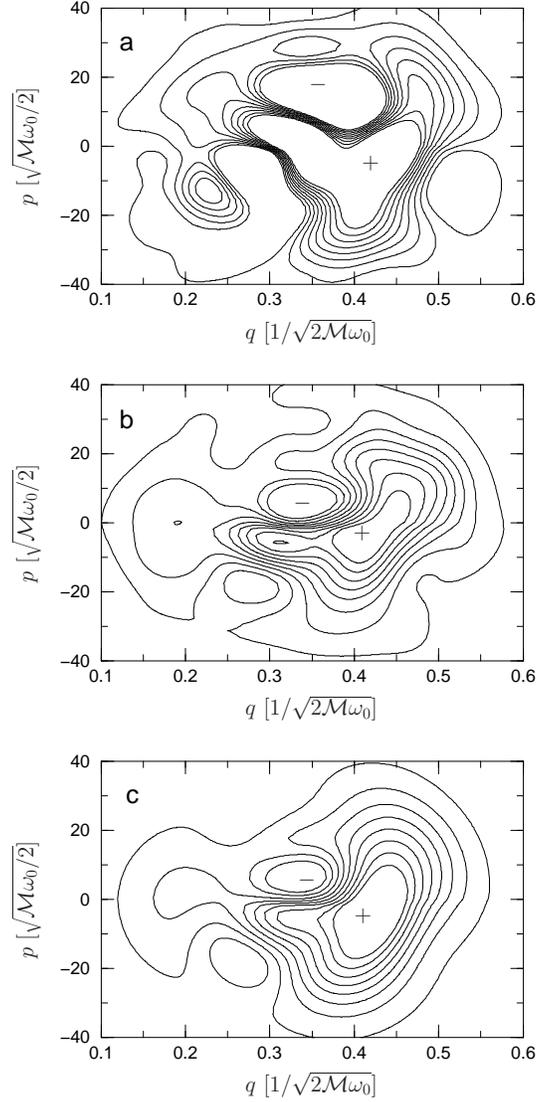}
\caption{
\label{fig:et-wigner-contour}
Contour plots of the Wigner representation of the RDM of one individual
trajectory (a), of the RDM recovered with 500 trajectories (b) and of the exact RDM (c). Regions with maximum are denoted with $+$, and regions with minimum with $-$.}
\end{figure}

\begin{figure}
\parbox[c]{\textwidth}{
\includegraphics[width=8.5cm,clip]{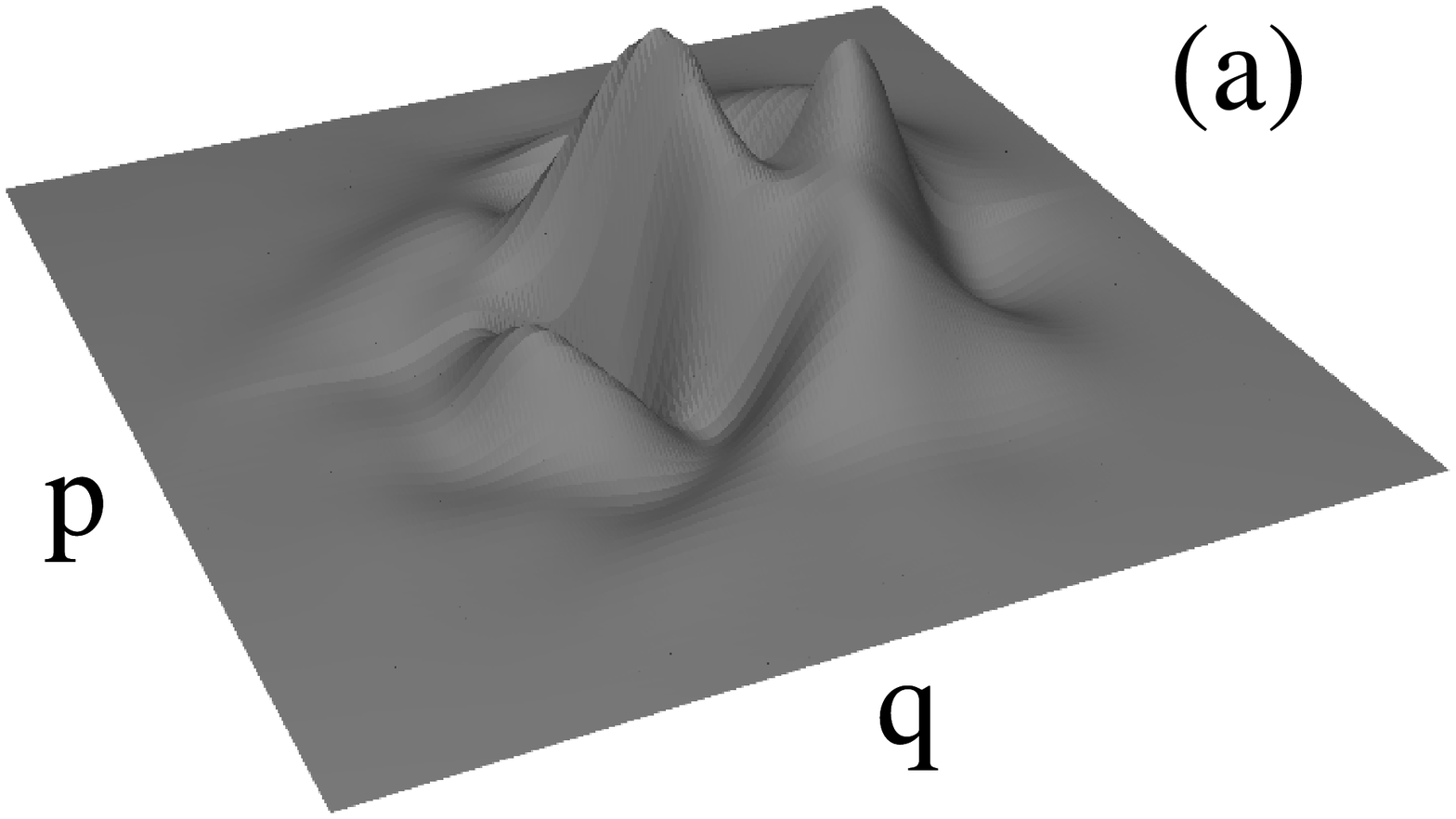}
\includegraphics[width=8.5cm,clip]{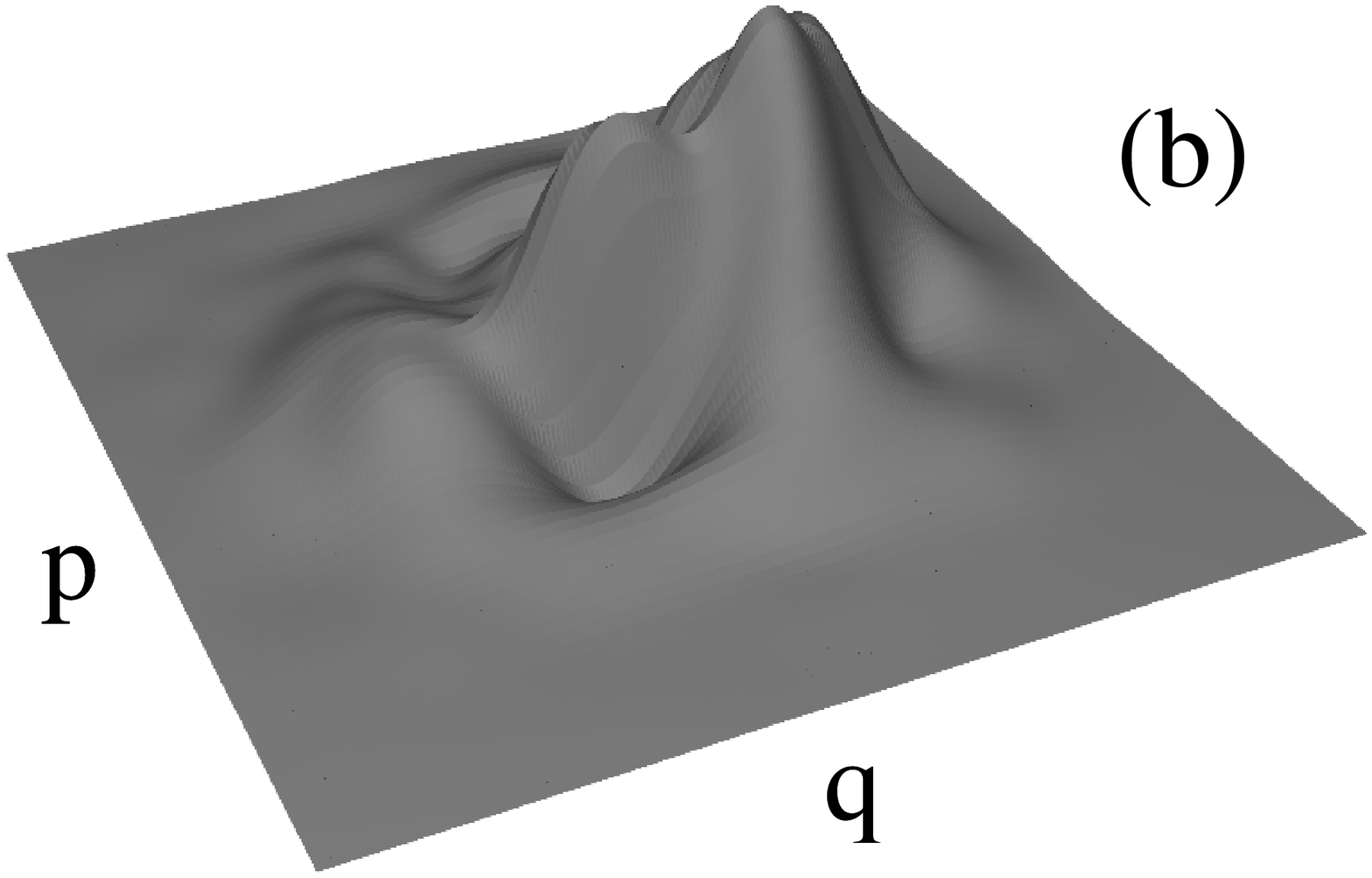}
\includegraphics[width=8.5cm,clip]{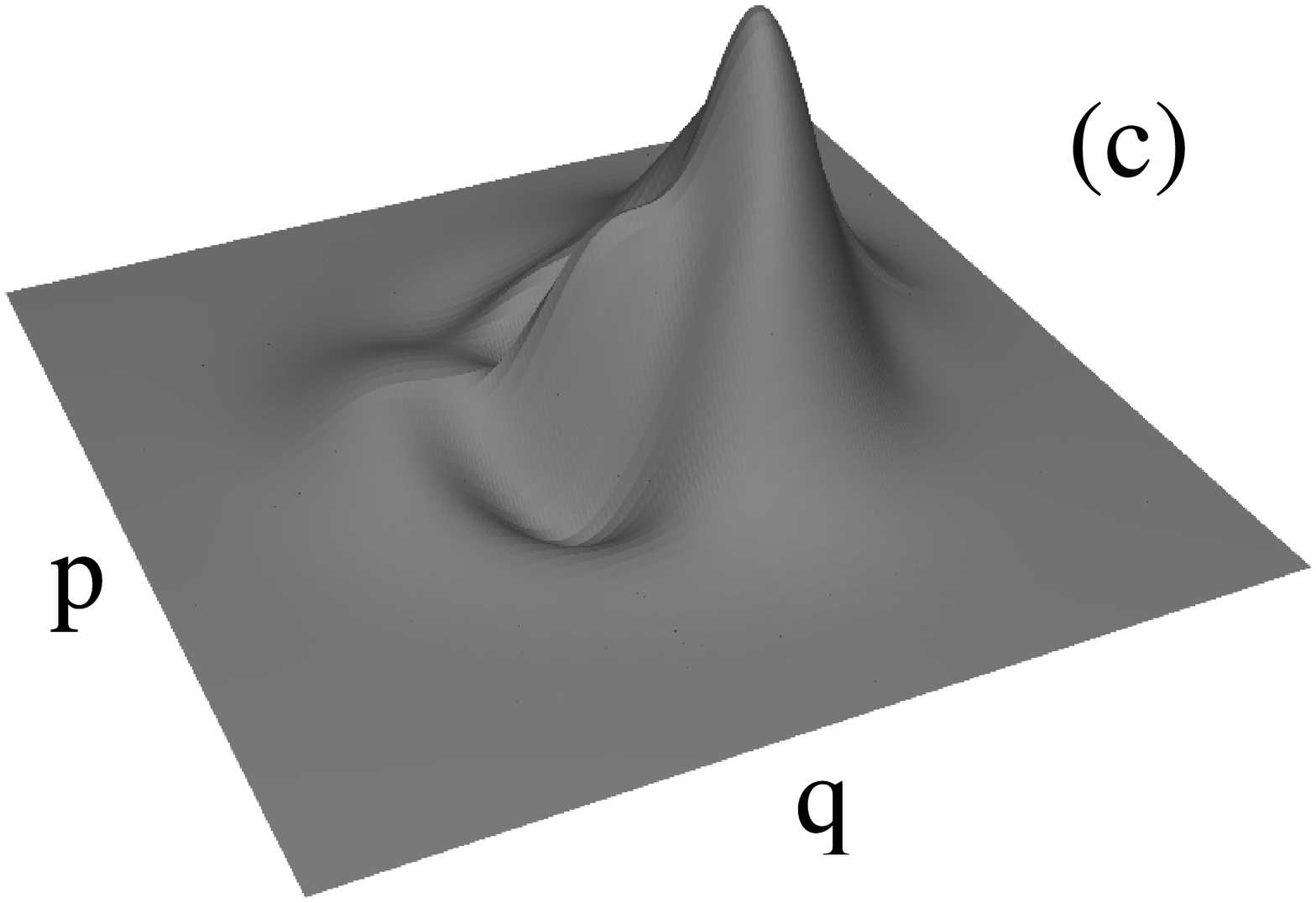}
}
\caption{
\label{fig:et-wigner-surface}
Three-dimensional plots of the Wigner representation of the RDM of one
individual trajectory (a), of the RDM recovered with 500 trajectories (b)
and of the exact RDM (c). The data shown here are the same as in
Fig.~\ref{fig:et-wigner-contour}. }
\end{figure}

\begin{figure}
\includegraphics[width=8.5cm]{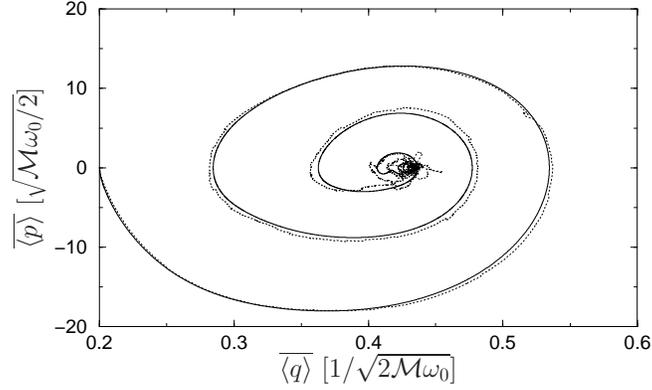}
\caption{
\label{fig:et-pq}
Dynamics of ET depicted as a path in phase space calculated by the exact
solution of the Redfield QME (solid line) and by averaging over 500
trajectories of the quantum jump method (dotted line). }
\end{figure}

Besides numerical efficiency, another advantage of the quantum trajectories
is the better insight into the quantum mechanisms underlying the overall
dynamics of the ensemble. Though it is impossible to give a direct physical
interpretation of every single trajectory one can extract information from
the ensemble statistics. As we can see in Fig.~\ref{fig:et-distrib} the
distribution of the individual expectation values of the population is skew
and comprises several maxima. One can better visualize the wave packet
dynamics in phase space using the Wigner representation of the RDM
\cite{hill84} as done in Figs.~\ref{fig:et-wigner-contour} and
\ref{fig:et-wigner-surface}. The evolution of the expectation values of the
momentum and the coordinate, which can be regarded as the center of mass of
the wave packet, is described by a path in phase space as shown in
Fig.~\ref{fig:et-pq}.

The wave packet starts off with zero momentum from the location of the
excited state. One can distinguish two stages of the ET dynamics. In the
first stage the wave packet splits into several parts and occupies the
whole accessible phase volume, i.e.,  it spreads (see
Figs.~\ref{fig:et-wigner-contour} b, c and \ref{fig:et-wigner-surface} b,
c). The motion of the principal part of the wave packet, which is seen as a
sharp peak in Fig.~\ref{fig:et-distrib}, implies coherent transfer of
population. The peak moves rapidly with time in an oscillatory fashion
while its amplitude decays as decoherence processes advance. After that all
parts of the wave packet coalesce to a single bell-shaped distribution. In
the second stage the wave packet continues to propagate slowly in phase
space while its maximum is approaching the equilibrium point at $\langle p
\rangle_{t\rightarrow\infty}$ and $\langle q \rangle_{t\rightarrow\infty}$.
This slow motion is seen as a small drift to the right beginning from the
central region of the spiral path (see Fig.~\ref{fig:et-pq}). It is due to
dissipative transfer mechanisms \cite{klei01} and is small for barrierless
potential configurations as in the present case. We note also that in the
time between the third population revival and the 10th vibrational period
(see Fig.~\ref{fig:et-pop}) one can recognize a crossover between the two
stages discussed above. Such a moment of time is shown in
Figs.~\ref{fig:et-distrib}, \ref{fig:et-wigner-contour} and
\ref{fig:et-wigner-surface}.

Unlike the trajectories considered in Ref.~\onlinecite{stru99} within the
quantum diffusion approach the wave packet of one individual trajectory in
our calculation spreads over the whole phase volume of the system
(Fig.~\ref{fig:et-wigner-contour} a). One reason for this discrepancy is
the different value used for $\hbar$. Generally, for sufficiently small
$\hbar$ the trajectories become classical states and virtually shrink to
points in phase space. However, the problem of their localization for small
$\hbar$ is non-trivial because the system may be almost classical but with
chaotic behavior.

\subsection{Electron transfer model with multiple reaction coordinates}
\label{subsec:et-2d}

Modeling the system-bath separation one has to minimize the degrees of
freedom in the relevant system and the system-bath coupling simultaneously.
The small polaron transformation has been used to effectively reduce the
system-bath coupling for a two-level spin-boson system \cite{poll96}. It
also has been shown that this approach can be extended for multi-level
systems \cite{poll96}. Alternatively, it is possible that one can
successively take strongly coupled degrees of freedom from the bath and put
them into the relevant-system part. This will make the effective
system-bath coupling smaller and hence the application of the Redfield
theory more reasonable. Multi-mode modeling of ET reactions, including
systems in the inverted region, has been done in
Refs.~\onlinecite{wolf95,wolf96}, and
\onlinecite{egor01} with similar argumentation. On the other
hand, there is experimental evidence for the participation of multiple
modes in the ET transition in some systems, such as oxazine-1 in
N,N-dimethylaniline \cite{wolf98} and betaine-30 in various solvents
\cite{kova01}. Correspondingly, the relevant part of the total ET system
can be modeled with a treatable small set of $R$ reaction coordinates
$\left\{ q_l \right\}$. For this purpose one may select a set of
representative harmonic normal modes from the molecule and from its
environment (e.g.\ the solvent or the crystal lattice). Since all normal
modes are decoupled one can use the single-mode operators to calculate the
matrix elements of the necessary operators in the diabatic basis $|i M_1
\dots M_R\rangle$. The Hamiltonian of each diabatic electronic state reads
\begin{eqnarray}
H_i=U_i
+\sum\limits_{l=1}^R
\left[
\omega_{i,l}(a_{i,l}^\dagger a_{i,l}+\frac12)
+\frac{\omega_{i,l}\Delta_{i,l}}{\sqrt2}(a_{i,l} + a_{i,l}^\dagger)
\right]~.
\label{eq:nd-hamil}
\end{eqnarray}
For $R$ reaction modes
Eq.~(\ref{eq:sb}) includes $R$ summation terms linear in each coordinate
$q_l$. The matrix element of $K_l$ reads
\begin{eqnarray}
\langle i M_1\dots M_R| K_l| j N_1 \dots N_R \rangle
=\frac{1}{\sqrt{2{\cal M}\omega_{i,l}}}
\delta_{ij}
\left(
\delta_{M_l+1,N_l}\sqrt{M_l+1}+\delta_{M_l-1,N_l}\sqrt{M_l}
\right)\prod\limits_{p\ne l}\delta_{M_p N_p}
\label{eq:nd-coord}
\end{eqnarray}
where ${\cal M}$ is the reduced mass of the relevant system.

The multi-mode model can be easily reduced to an effective single-mode
model by means of an orthogonal transformation \cite{obri72,taka78} of the Hamiltonian $H^{(0)}$. For two diabatic states with equal curvatures one can drop the electronic index. Denoting the relative displacement by
$\Delta_l$ the transformation has the form \cite{taka78}
\begin{eqnarray}
B_k=\sum\limits_{l=1}^R V_{kl} a_l~,
\label{eq:u-trans}
\end{eqnarray}
with
\begin{eqnarray}
V_{kl}=\frac{\Delta_l\omega_l}{\sqrt2 (\omega_l - \Omega_k)D_k}~,\quad
D_k^2=\sum\limits_{l=1}^R\frac{\Delta_l\omega_l}{\sqrt2 (\omega_l - \Omega_k)^2}, \quad\textrm{for}\quad k\ne 0
\label{eq:u-trans-1}
\end{eqnarray}
and
\begin{eqnarray}
V_{0l}=\frac{\Delta_l\omega_l}{(\Delta_0\Omega_0)^2},\quad
(\Delta_0\Omega_0)^2=\sum\limits_{l=1}^R(\Delta_l\omega_l)^2,\quad
\Omega_0=\frac{1}{(\Delta_0\Omega_0)^2}\sum\limits_{l=1}^R\omega_l^3\Delta_l^2~.
\label{eq:u-trans-2}
\end{eqnarray}
For $k\ne0$ the new mode frequencies $\Omega_k$ are the roots of the equation
\begin{eqnarray}
\sum\limits_{l=1}^R\frac{\Delta_l\omega_l}{\omega_l - \Omega_k}=0~.
\label{eq:u-trans-3}
\end{eqnarray}
After this transformation the multi-mode Hamiltonian is cast into the form (\ref{eq:Hamiltonian}) where
\begin{eqnarray}
H_{\rm S}&=& U_0+\Omega_0 B_0^\dagger B_0+\frac{\Omega_0 \Delta_0}{\sqrt2}
(B_0+B_0^\dagger)\\
H_{\rm B}&=&\sum\limits_{k=1}^{R-1}\Omega_k B_k^\dagger B_k\\
H_{\rm SB}&=&\sum\limits_{k=1}^{R-1}\frac{\Delta_0\Omega_0}{D_k} (B_0^\dagger B_k + B_k^\dagger B_0)~.
\label{eq:after-u}
\end{eqnarray}
One can see that all $R$ normal modes are transformed to a finite bath with
$R-1$ modes with frequencies $\Omega_k$. A new effective mode with
frequency $\Omega_0$ and displacement $\Delta_0$ is created which is
bilinearly coupled to the new bath modes. From Eq.~(\ref{eq:u-trans-2}) it 
follows that $\Delta_0\Omega_0<\sum_l\Delta_l\omega_l$ for positive
coordinate displacements, i.e.\ the vibronic coupling of the new effective
mode is smaller than the total vibronic coupling of the initial normal
modes. Thus, a reduction from a multi-mode to a single-mode model extends
the bath and enlarges the system-bath coupling. This reduction is unique.
On the other hand, the addition of a bath mode to the relevant system is
not unique. It depends on the choice of a certain bath mode. Nevertheless,
it always reduces the system-bath coupling and enlarges the total vibronic
coupling of the relevant system.

\begin{figure}
\includegraphics[width=8.5cm]{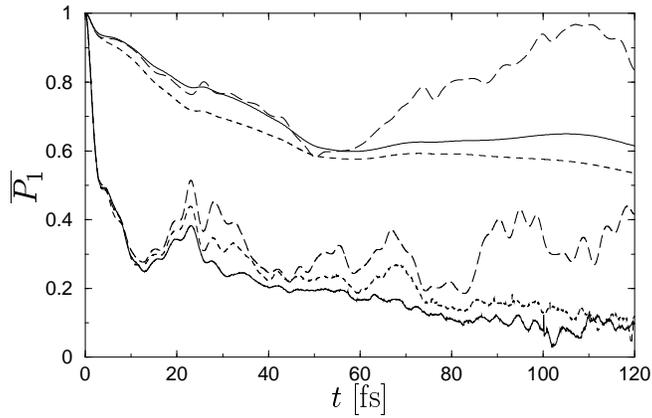}
\caption{
\label{fig:et-2d-1d}
Population dynamics of the excited state in the effective one-mode (thick
lines) and the two-mode (thin lines) models for ET. Coherent dynamics are
denoted by long dashed lines, dynamics with dissipation in DDA by dashed
lines, and the Redfield dynamics by solid lines. The two-mode model is
solved with the new stochastic method with 5000 trajectories. }
\end{figure}

In the following, a two-mode ET model will be considered. The frequencies
of the two modes were chosen $0.07$ and $0.18$ eV, and the reorganization
energies $0.33$ and $0.82$ eV, respectively. These frequencies correspond
to internal molecular modes in real ET systems. The effective mode was
calculated using Eq.~(\ref{eq:u-trans-2}). The change of the free energy
was taken to be $U^0_2-U^0_1=-0.2$ eV and the electronic coupling between
the diabatic electronic states $v_{12}=0.1$ eV. Again, a harmonic bath with
an Ohmic spectral density was considered with damping rate $\Gamma=0.007$
eV and temperature $295$ K. 16 levels for each mode gave a good
convergence. Initially, the lowest vibrational level of the excited state
is populated (Fig.~\ref{fig:et-2d-1d}). The coherent dynamics of the
excited state of the effective mode model yields small population in the
ground state. This is followed by an almost complete revival at about 110
fs. In contrast, the two-mode model exhibits an ultrafast coherent
population transfer to the ground state. Recurrences of population in the
excited state appear many times within 120 fs but their yield does not
exceed 50 \%. As the two-mode system can be regarded as a conservative
two-particle system its complete recurrence period (Poincare's cycle) is
much longer than that for the single-mode system. In a way the isolated
two-mode system shows a relaxation behavior that is typical for open
systems.

Turning on the dissipation leads to irreversible transfer to the lower
state. However, the picture does not change qualitatively as the
dissipative transfer mechanism does not contribute significantly at the early
times shown here. As already found \cite{egor01,klei01} and seen in
Fig.~\ref{fig:et-2d-1d} the DDA has also no serious influence on the early
dynamics. As the DDA has been recently well studied we have put the
focus here on the ultrafast initial stage where the difference between  the single-mode model and the two-mode model is best characterized.

\section{Efficiency and stability}
\label{sec:efficiency}

To estimate the convergence and hence the stability of the proposed scheme
one needs an appropriate measure for the convergence.  Unraveling the QME
one aims to calculate the time evolution of an observable, e.g.\ the
population $\overline{P_1}$, performing an average over $N_{\textrm{s}}$
single trajectories. At time $t_j$ the average reads
\begin{equation}
\overline{P_1}(t_j,N_{\textrm{s}})=\frac{1}{N_{\textrm{s}}}
\sum\limits_{i=1}^{N_{\textrm{s}}} 
\left[
\langle\psi_i(t_j)|1\rangle\langle1|\phi_i(t_j)\rangle+\textrm{c.c.}
\right]
\label{eq:partial-average}~.
\end{equation}
As a convergence measure we introduce the quantity
\begin{equation}
\varepsilon^2(N_{\textrm{s}},k)=\frac{1}{N_t}\sum\limits_{j=1}^{N_t} 
\left[
\overline{P_1}(t_j,N_{\textrm{s}})-\overline{P_1}(t_j,N_{\textrm{s}}-k)
\right]^2
\label{eq:varepsilon},
\end{equation}
where $N_t$ is the total number of propagation time steps. In this
convergence measure we use as reference the average performed over
$N_{\textrm{s}}-k$ trajectories. For convenience the increment $k$ can be
chosen to be the number of computing nodes in a parallel implementation of
the stochastic algorithm. One can easily see that $\varepsilon$ vanishes
for  large $N_{\textrm{s}}$ if both terms in the sum converge to
$\overline{P_1}(t_j,N_{\textrm{s}}\rightarrow\infty)$. If these two terms
diverge with increasing $N_{\textrm{s}}$ the scheme is instable. Therefore
$\varepsilon$ is sufficient for estimating the convergence and stability
limits. The measure $\varepsilon$ used here is very similar to the absolute
error measure $\beta$ for the unraveling schemes which has been used for
studying standard first- and higher-order unraveling schemes \cite{stei95}.
The only difference between them is the reference calculation -- for
calculating the error $\beta$ one considers the exact solution produced
numerically by some direct propagator or analytically. It was found
\cite{stei95} that for a small number of samples $N_{\textrm{s}}$ the error
measure $\beta$ is mainly statistical due to the finite sample size. For
large $N_{\textrm{s}}$ the error due to the finite time step starts to
dominate.  Since both terms in the sum in Eq.~(\ref{eq:varepsilon}) carry
the same time-step error it will cancel out in the error measure
$\varepsilon$. Thus our convergence measure is  a criterion for
the statistical error of the stochastic scheme.

\begin{figure}
\includegraphics[width=8.5cm]{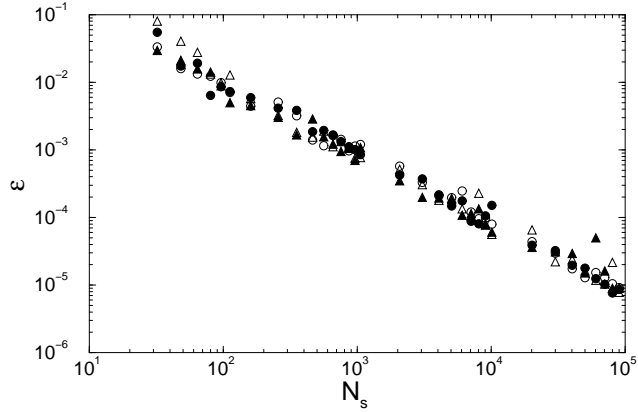}
\caption{
\label{fig:convergence}
Convergence behavior of the proposed stochastic unraveling scheme for
$\Gamma=0.01\omega_0$ (opaque), $\Gamma=0.1\omega_0$ (filled), $\delta t=1$ (circles), and $\delta t=10$ (triangles).}
\end{figure}

Considering the efficiency of the stochastic method it is convenient to
look at the relation between numerical effort and achieved convergence.
When the DDA is invoked together with the RWA in the system-bath coupling
one can obtain the Lindblad form (\ref{eq:lindblad-form}) that can be
solved by the standard quantum jump method
\cite{dali92,gard92,gisi92,garr94,stei95,plen98} as well as by the present
one. In this way the numerical performance of both methods can be compared.
Figure \ref{fig:convergence} shows the behavior of $\varepsilon$ for $k=8$
with increasing number of trajectories for the ET example with one reaction
coordinate. Within Redfield theory the variation of $\Gamma$ towards large
system-bath coupling values is of limited validity. So the Redfield QME
does not allow to verify the new scheme in the strong coupling regime.
However, we performed the calculations for two values for $\Gamma$
($0.1\omega_0$ and $0.01\omega_0$) that are supposed to belong to the weak
coupling regime. In addition, the computation was performed for two
propagation time steps $\delta t$. It was found that for $\Gamma \delta
t\approx 0.1$ the proposed scheme becomes unstable. As can be seen in
Fig.~\ref{fig:convergence} decreasing $\Gamma \delta t$ has no influence on
the convergence $\varepsilon$.  The slope of $\varepsilon$ in the double
logarithmic scale ($d\log\varepsilon/d\log N_{\textrm{s}}$) in Fig.~\ref{fig:convergence} is $-1$, i.e.\ the scheme
converges as $1/N_{\textrm{s}}$.  When simulating other, physically
different systems we expect the convergence behavior to be not too
different from the example studied here. At least the proportionality to
$1/N_{\textrm{s}}$ will stay unchanged. The only difference can be the
intercept of $\varepsilon$ (i.e. the value of
log epsilon for $\log N_{\textrm{s}}=0$) which may have some physical reasoning. We
expect some change in the instability limit for $\Gamma \delta t \approx 0.1$
with the type of system studied.  This topic has to be explored in future
studies, especially for QMEs which allow larger variation of the
system-bath coupling.

\begin{figure}
\includegraphics[width=8.5cm]{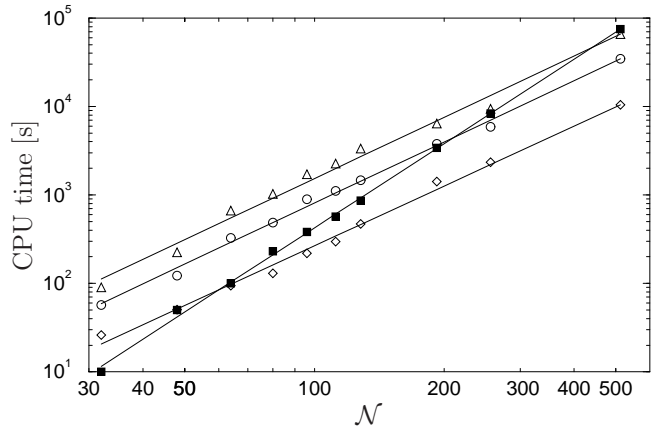}
\caption{
\label{fig:effort}
Numerical effort of the new unraveling scheme with 500 trajectories for
Eq.~(\ref{eq:pf-form}) (circles), the standard quantum jump method for
Eq.~(\ref{eq:lindblad-form}) (rhombs), and the new scheme for
Eq.~(\ref{eq:lindblad-form}) (triangles) shown for the model of
one-dimensional ET. The short iterative Arnoldi method (filled squares) is
shown as reference solving Eq.~(\ref{eq:pf-form}). The data points at
${\cal N}=512$ are extrapolated. }
\end{figure}

Now let us discuss the dependence on the basis size. It is well known that
the numerical effort for solving QMEs with the quantum jump method and with
direct propagators scales quadratically and cubically, respectively, with
the basis size ${\cal N}$. Thus, for both high accuracy and lower numerical
expense stochastic methods would only be preferred over direct propagators
if the number of trajectories $N_{\textrm{s}}$ is much smaller than ${\cal
  N}$. We shall examine the scaling behavior by solving the one-dimensional
ET model with the use of the present stochastic scheme for both the
Redfield QME (\ref{eq:pf-form}) and the Lindblad QME
(\ref{eq:lindblad-form}) as well as the standard scheme for the Lindblad
QME (\ref{eq:lindblad-form}). All runs were performed with increasing basis
size and compared to a direct propagation. For this purpose we choose the
short iterative Arnoldi propagator \cite{poll94,poll96,kond01} in a Krylov
space of dimension 12.  As expected, the numerical expense for a few
trajectories is much smaller than for the direct propagator.  But for
converged results one needs much more that one trajectory. When the number
of trajectories necessary for the complete convergence is greater than
${\cal N}$ one has to make a trade-off between the accuracy achieved with
the stochastic method and the numerical effort.  For the comparison we
choose as an example $N_{\textrm{s}}=500$ for which the convergence is not
yet complete. In Fig.~\ref{fig:et-pop} we have seen that this sample size
already yields qualitatively the same result as the direct propagator. For
very accurate calculations one needs a much larger number of trajectories.
The crossing point in Fig.~\ref{fig:effort} shows that for ${\cal N}
\gtrsim 212$ the new stochastic algorithm should be preferred. In addition
one gets a benefit from the efficient parallel implementation of the
stochastic algorithm which is of great practical use especially when one is
asking for fast preliminary results for the dynamics of large systems. The
slope of the lines in Fig.~\ref{fig:effort} reflecting the scaling of the
numerical effort for the stochastic methods is approximately $2.3$ and for
the direct propagation $3.2$. These scalings are larger than the
theoretical estimates of 2 and 3, respectively, because the true complexity
of the algorithm is not a simple power function but a polynomial function
of ${\cal N}$ due to array operations of lower order.  This difference has
to disappear for very large ${\cal N}$.

In the following the intercept of the curves for the numerical effort
in Fig.~\ref{fig:effort}, i.e.\ the CPU-time extrapolated to ${\cal N}=1$, is examined.  The ratio of the intercepts between the proposed stochastic
scheme for the Redfield QME and the standard scheme for the Lindblad QME is
found to be $2.4$ (see Fig.~\ref{fig:effort}). The respective ratio
between the proposed stochastic scheme for the Redfield QME and the same
scheme for the Lindblad QME is $4.4$. This can be interpreted as follows.
Neglecting the operations of lower order the intercept in a double
logarithmic plot like in Fig.~\ref{fig:effort} has to be proportional to
the number of matrix-vector multiplications for one time step, i.e.\ the
intercept holds to some extent information specific for each stochastic
scheme. Considering only the deterministic part of the implemented
algorithm (see the Appendix for details) each time step requires $4M+9$
matrix-vector multiplications (for $M$ dissipative channels, $4\times{}M$
operations for computing the jump rates Eq.~(\ref{eq:jump-rates}), eight
operations for time propagation and one for calculating the population)
versus only $5$ for the standard jump algorithm (four operations for time
propagation and one for calculating the population). Nevertheless, the
values found from the simulation (Fig.~\ref{fig:effort}), $12/5$ and
$22/5$, deviate from the estimates $13/5$ and $17/5$, respectively, due to
systematic effects like lower-order operations for not very large ${\cal
  N}$ as well as due to stochastic effects like the varying number of quantum
jumps. To summarize, for the Lindblad QME the standard algorithm is the
method of choice, while for the Redfield QME, where one cannot apply the
standard method, one has to cope with the larger numerical expense. The
latter can be significantly reduced when high accuracy is not necessary.

\section{Conclusion}
\label{sec:conclusion}
In this paper the stochastic unraveling technique was extended for
non-Lindblad QMEs. This progress became possible with the use of the
wave-function pair in the doubled Hilbert space and the derivation of
stable, almost normalized SSEs. An efficient first-order quantum jump
algorithm was proposed. The efficiency is determined by the behavior of the
norm of every single trajectory. In this sense the jump rates were used as
parameters to influence the efficiency.

Occurrence of negative population for single trajectories is by no means a
problem of the proposed unraveling scheme. Rather it is related to the fact
that the QMEs (\ref{eq:pf-form}, \ref{eq:qme}) do not preserve the RDM
positive semi-definite. It is known \cite{gasp99a} that the negative
eigenvalues of the RDM in the Redfield theory arise from the inconsistency
between the initial RDM and the bath state, i.e.\ due to neglected initial
correlations in the Born-Markov approximation. A satisfactory resolution of
this problem is the slippage of the initial conditions as derived by
Gaspard \emph{et al.}\ \cite{gasp99a}. In this method the so called slippage
superoperator takes into account the short-time bath correlations. Applied
to the initial RDM it introduces the necessary correlations into the
initial state. This manipulation of the initial state ensures propagation
of a positive semidefinite RDM at any further moment of time.

The method proposed was successfully tested for the Redfield QME for the
damped harmonic oscillator and two ET models. It was shown that the scheme
allows for more efficient quantum dynamical simulations of large systems.
The most important benefit of the method is that it can be applied in a
straightforward manner since the SSEs and the expressions of the jump rates
are formally the same for other models of system and bath. Therefore, a
potential use of the proposed method can be made in simulations involving any kind of non-Markovian QMEs provided that they are in a time-local form like in the time-convolutionless formalism \cite{breu99} as well as in methods using auxiliary density matrices to include the memory effects \cite{meie99}.

Each individual trajectory occupies nearly the whole volume in phase space
accessible for the system. It would be of possible interest to see whether
the phase-space volume of a single trajectory shrinks with decreasing
$\hbar$ and to study the classical behavior of the ET system.


\acknowledgments

The authors acknowledge financial support from the DFG.

\begin{appendix}
\section*{Appendix: The numerical algorithm}
\label{sec:mc-nolindblad-alg}

This algorithm gives the numerical solution of the SSEs (\ref{eq:sse1_a})
and (\ref{eq:sse1_b}). The initial wave functions $|\psi_{i,s}(0)\rangle$ and
$|\phi_{i,s}(0)\rangle$ are constructed so that
\begin{equation}
\rho_{\rm S}(0)
=
\sum\limits_{i=1}^{N_e} 
w_i
\frac{1}{N_{\rm s}}
\sum\limits_{s=1}^{N_{\rm s}}
\left[
|\phi_{i,s}(0)\rangle
\langle \psi_{i,s}(0)|
+
|\psi_{i,s}(0)\rangle
\langle \phi_{i,s}(0)|
\right]~.
\label{eq:2d-recover-0}
\end{equation}
Here $N_e$ is the number of non-zero eigenvalues $w_i$ of the initial
density matrix $\rho_{\rm s}(0)$ and $N_{\rm s}$ the number of trajectories
corresponding to each eigenvalue $w_i$. The wave functions are propagated jointly (as pairs) as follows starting with $t=0$.
\begin{enumerate}
\item\label{item:2d-store} store/send $|\psi_{i,s}(t)\rangle$ and
  $|\phi_{i,s}(t)\rangle$ for averaging;
\item\label{item:2d-jump-rate} calculate the rates $p_k^1$ and $p_k^2$
  according to Eqs.~(\ref{eq:jump-rates_a}) and (\ref{eq:jump-rates_b}).
        \item generate a random number $\epsilon\in(0,1)$;
        \item if $\epsilon > dt\sum_k (p_k^1+p_k^2)$ then\\
        propagate $|\psi_{i,s}(t)\rangle$ and $|\phi_{i,s}(t)\rangle$:
        \begin{itemize}
        \item[$\ast$] find $|\psi_{i,s}(t+dt)\rangle$ and  $|\phi_{i,s}(t+dt)\rangle$ solving\\
          $d|\psi_{i,s}(t)\rangle/dt= A |\psi_{i,s}(t)\rangle$ and
          $d|\phi_{i,s}(t)\rangle/dt= A |\phi_{i,s}(t)\rangle$,  respectively
        \item[$\ast$] set $t=t+dt$
        \item[$\ast$] go to step \ref{item:2d-store}
        \end{itemize}
\vspace{-2ex}
        else
        \begin{itemize}
        \item[$\ast$] if $\epsilon \le dt\sum_k p_k^1$ then 
        \begin{itemize}
        \item[--] jump with probability $p_k^1 dt$:\\
          $|\psi_{i,s}(t)\rangle \rightarrow E_k |\psi_{i,s}(t)\rangle/\sqrt{p_k^1}$ and $|\phi_{i,s}(t)\rangle \rightarrow C_k |\phi_{i,s}(t)\rangle/\sqrt{p_k^1}$\\
        \end{itemize}
\vspace*{-3ex}
        else
        \begin{itemize}
        \item[--] jump with probability $p_k^2 dt$:\\
          $|\psi_{i,s}(t)\rangle \rightarrow C_k |\psi_{i,s}(t)\rangle/\sqrt{p_k^2}$ and $|\phi_{i,s}(t)\rangle \rightarrow E_k |\phi_{i,s}(t)\rangle/\sqrt{p_k^2}$\\
        \end{itemize}
\vspace*{-3ex}
        \item[$\ast$] go to step \ref{item:2d-jump-rate}
        \end{itemize}
\end{enumerate}
The deterministic propagation of the wave functions in this work was
performed with the use of a forth-order Runge-Kutta method \cite{pres92}.
Accordingly, one time step requires four matrix-vector multiplication for
each wave function. The ensemble-averaged expectation value of an
observable $A$ is calculated as
\begin{equation}
\overline{\langle A(t) \rangle}
=
\sum\limits_{i=1}^{N_e} 
w_i 
\frac{1}{N_{\rm s}}
\sum\limits_{s=1}^{N_{\rm s}}
\left[
\langle \psi_{i,s}(t)|
A
|\phi_{i,s}(t)\rangle
+
\textrm{c.c.}
\right]
~.
\label{eq:2d-recover-observ}
\end{equation}

The method can be parallelized using MPI \cite{mpif94}. In such an
implementation every single stochastic trajectory is propagated by a
different process. Only the averaging operation
(\ref{eq:2d-recover-observ}) is done at certain times by means of
collective communications. In this way the task can be efficiently
distributed on a cluster of PCs.

\end{appendix}

\newpage

\end{document}